\documentclass[12pt,a4paper]{article}

\makeatletter
\@addtoreset{equation}{section}
\makeatother

\newcommand{\bsa}{$\mbox{M5} \mbox{--} \overline{\mbox{M5}}$}
\newcommand{\bsb}{$\mbox{MKK} \mbox{--} \overline{\mbox{MKK}}$}
\newcommand{\bsc}{$\mbox{M9} \mbox{--} \overline{\mbox{M9}}$}
\newcommand{\bsd}{$\mbox{M2} \mbox{--} \overline{\mbox{M2}}$}

\begin{document}
\begin{flushright}
\footnotesize
\footnotesize
CERN-TH-99-331\\
Imperial/TP/99-0/8\\
November, $1999$
\normalsize
\end{flushright}

\begin{center}

\vspace{.8cm}
{\LARGE {\bf Type II Branes from Brane-Antibrane in M-theory}}

\vspace{1cm}


{\bf Laurent Houart}

\vspace{.1cm}

{
{\it Theoretical Physics Group\\
Blackett Laboratory,
Imperial College \\
London SW7 2BZ, UK}\\
{\tt l.houart@ic.ac.uk}
}

\vspace{.3cm}

{and}

\vspace{.3cm}

{\bf Yolanda Lozano}

\vspace{.1cm}

{
{\it Theory Division, CERN\\
1211 Gen\`eve 23, Switzerland}\\
{\tt yolanda.lozano@cern.ch}
}

\vspace{.4cm}

\vspace{1cm}


{\bf Abstract}

\end{center}
\begin{quotation}

\small

We discuss in a systematic way all the possible realisations of branes
of M and type II theories as topological solitons of a brane-antibrane
system. The classification of all the possibilites, consistent with the
structure of the theory, is achieved by studying the Wess-Zumino terms 
in the worldvolume effective actions of the branes of M-theory and 
their reductions.

\end{quotation}

\vspace{1cm}

\newpage

\pagestyle{plain}

\section{Introduction}

During the past couple of years, significant progress has been made
in our understanding of M-theory beyond the BPS configurations.
These advances have already permitted to test the web of dualities 
relating the different phases of M-theory  (the different superstring 
theories) on some of the  non-BPS states of the spectrum,  
and a beautiful outlook on the interplays between BPS branes
and non-BPS branes has been given in some cases (see
\cite{reva}-\cite{revc} for reviews and references therein) along with
an elegant mathematical formulation \cite{kteo,hora2}. 
For example, a BPS  Dp-brane of type II theory may be viewed as coming
from a non-BPS system given by a D(p+2), anti D(p+2) pair \cite{senb},
\cite{sena}. 
The instability of this non-BPS configuration manifests itself in a 
complex tachyonic mode of the open string stretched between the pair. 
When the pair coincides, the
tachyon rolls down to a true vacuum and condenses leading to a stable  
vortex-like configuration, and the resulting object is
a BPS Dp-brane. Recently, it has been argued \cite{Yi} that 
a similar mechanism could also produce fundamental strings. 
Namely, a fundamental string in the Type II theory
could be described as a bound state of any pairs of stable Dp, anti-Dp
branes where the tachyon of a D(p-2)-brane stretched between them 
condenses. 
In \cite{Yi}, the author considers in some detail the case $p=4$, 
i.e. the condensation of a D2-brane  stretched between a D4, anti-D4
pair. Since the tachyonic condensing charged object is in this case 
extended (a tachyonic worldvolume string), there are no direct ways  
to describe quantitatively  this type of mechanism.
Nevertheless, the existence of this process can be deduced from the 
following three steps consideration \cite{Yi}. 
One first notices that the usual process \cite{sena} of creation of a  
D2-brane in terms of the condensation of a fundamental string should have 
a description  in M-theory in terms of creation of an M2-brane 
by condensation of a stretched M2 between a pair of M5, anti-M5 branes.
Then one identifies in the worldvolume effective action of the brane
anti-brane pair
the Wess-Zumino term responsible for this process. Finally upon
dimensional reduction one finds two Wess-Zumino terms, one describing
the usual realisation of the D2 as a soliton, the other describing the
fundamental string.

In this paper, we propose to extend in a systematic way this kind
of consideration. We study all the possible realisations of branes
of M and type II theories as topological solitons of a brane-antibrane
system by looking at the Wess-Zumino terms of the worldvolume effective 
actions of the different brane anti-brane pairs in M-theory.

The paper is organised as follows. In section 2 we study the {\bsa}
system, reviewing the work of Yi \cite{Yi}. Section 3 discusses the
{\bsb} system, and in section 4 we study the {\bsc} case. 
Section 5 considers all brane-antibrane systems in M-theory in which 
M-waves are involved, in particular the {\bsd} case.
Finally, section 6 analyses type IIB branes from the same kind of
brane- antibrane systems.
The last section contains a summary  and some discussions.

\section {The {\bsa} system}

In this section we briefly review the results of ref.\cite{Yi}.
We start with the {\bsa} system and analyse the process
of annihilation of a pair of M5, anti-M5 branes
in terms of the tachyonic condensation of an M2-brane stretched between
this pair. 
In particular, one can identify the following coupling \cite{chs} 
in the M5-brane, $\overline{\mbox{M5}}$-brane worldvolume action:
\begin{equation}
\label{Mcoupling}
\int_{R^{5+1}}{\hat C}\wedge d{\hat a}^{(2)}\, ,
\end{equation}

\noindent as the one describing the emergence of an M2-brane soliton 
when the stretched tachyonic M2-brane condenses \cite{Yi}. Here
${\hat C}$ is the 3-form of eleven dimensional supergravity
and ${\hat a}^{(2)}$ the worldvolume 2-form present in the action of the
M5-brane\footnote{ Hats on target space fields indicate that they are 
11-dimensional. We use hats as well for the worldvolume fields of
branes in 11 dimensions.}. This 2-form 
is self-dual for a single M5-brane, however
for an M5, anti-M5 pair it is unrestricted \cite{Yi}, 
given that in the anti-M5
brane effective action it is anti-self-dual and both contributions
are combined to describe the coinciding M5, 
anti-M5 pair\footnote{The complete WZ
term including the coupling of the complex tachyonic field has been
constructed in \cite{KW} for D brane anti-D brane systems.}. 
The topologically non-trivial tachyonic condensation 
of an M2  is though \cite{Yi} to be accompanied by a localised magnetic flux 
$d{\hat a}^{(2)}$. Integrating over the flux on a transverse $R^3$ one 
finds:

\begin{equation}
\int_{R^{2+1}}{\hat C}\, ,
\end{equation}

\noindent which means that the condensation of the tachyonic mode of
the M2  gives rise to the annihilation of the M5 anti-M5 pair into an 
M2-brane, since this is the way the M2-brane couples, minimally, to ${\hat C}$.

The dimensional reduction of the stretched M2-brane between the two
M5-branes, along an M5-brane worldvolume direction,
gives a fundamental string stretched between a D4 and an
anti-D4 branes, if the reduction takes place along the M2-brane, 
or a D2-brane stretched between two D4, anti-D4 branes, 
if the reduction
takes place along a direction transverse to the M2-brane. These two
processes are described by the worldvolume reduction of 
(\ref{Mcoupling})\footnote{We ignore all numerical prefactors.}:

\begin{equation}
\label{reda}
\int_{R^{5+1}}{\hat C}\wedge d{\hat a}^{(2)} \sim \int_{R^{4+1}}
C^{(3)}\wedge db^{(1)}+\int_{R^{4+1}}B^{(2)}\wedge da^{(2)}
\end{equation}

\noindent where:

\begin{equation}
{\hat a}^{(2)}_{\mu 5}=b^{(1)}_\mu\, , \qquad
{\hat a}^{(2)}_{\mu\nu}=a^{(2)}_{\mu\nu}\, , \qquad \mu=0,1,\dots,4\, ,
\end{equation}  

\noindent and $C^{(3)}$ denotes the RR 3-form and $B^{(2)}$ the NS-NS
2-form of the Type IIA theory. The first term describes a stretched
fundamental string, coupled to $b^{(1)}$, and the second a stretched
D2-brane, coupled to $a^{(2)}$.
Considering the first term, integration over the localised magnetic 
flux $db^{(1)}$, which accompanies 
the topologically non-trivial condensation of the tachyon of the string 
\cite{sena}, along a transverse  $R^2$ gives the coupling:

\begin{equation}
\int_{R^{2+1}}C^{(3)}
\end{equation}

\noindent which is the way the RR 3-form couples to the D2-brane.
Therefore, one recovers  the fact \cite{sena} that the condensation 
of the stretched fundamental string, coupled to $b^{(1)}$, gives a 
solitonic D2-brane. Reversing the logic, the well-established process
of creation of a D2-brane legitimates by oxidation the process of M2
creation described above. 

On the other hand, a similar argument \cite{Yi} on the second term 
of (\ref{reda}) shows that the mechanism of tachyonic condensation of
the stretched  D2-brane, coupled to $a^{(2)}$ , upon integration 
over the localised flux $da^{(2)}$ on a transverse $R^3$  
(flux which should accompany 
this topologically non-trivial process) gives:

\begin{equation}
\int_{R^{1+1}}B^{(2)}\, ,
\end{equation}

\noindent describing a solitonic fundamental string.

One can describe as well the process in which a D2-brane stretched
between a NS5, anti-NS5 pair of branes condenses, giving rise to a
D2-brane soliton. The corresponding coupling is given by the reduction
of (\ref{Mcoupling}) along a direction perpendicular to the M5-branes
worldvolume:

\begin{equation}
\int_{R^{5+1}}C^{(3)}\wedge da^{(2)}\, .
\end{equation}

\noindent Here the integration over the localised flux $da^{(2)}$ on 
a transverse 
$R^3$ gives the minimal coupling of a D2-brane:

\begin{equation}
\int_{R^{2+1}}C^{(3)}\, .
\end{equation}

\section{The {\bsb} system}

Following the same reasoning as in the previous section
we now analyse the different possible processes starting with the
{\bsb} system by looking at the relevant terms in the worldvolume
effective action of the Kaluza-Klein monopole.

The worldvolume effective action of the M-theory Kaluza-Klein monopole
was constructed in \cite{BJO,BEL}. The existence of the Taub-NUT 
direction in the space transverse to the monopole
is implemented at the level of the effective action by 
introducing a Killing isometry which is gauged in the worldvolume. 
Then the target space fields must couple in the worldvolume
with covariant derivatives of the embedding scalars, or through 
contraction with the Killing vector. The Kaluza-Klein monopole is charged
with respect to an 8-form, which is the electric-magnetic dual of the
Killing vector considered as a 1-form. 
This field is itself contracted with the Killing vector, giving a 7-form
minimally coupled to the 7 dimensional worldvolume of the monopole.

The worldvolume effective action of the monopole contains the following
term \cite{BEL}:
\begin{equation}
\label{M2deM5}
\int_{R^{6+1}} i_{\hat k}{\hat {\tilde C}}\wedge d{\hat b}^{(1)}\, ,
\end{equation}

\noindent where ${\hat {\tilde C}}$ denotes the 6-form of eleven 
dimensional supergravity, ${\hat k}$ is the Killing vector, with
$(i_{{\hat k}}{\hat {\tilde C}})_{{\hat \mu}_1\dots {\hat \mu}_5}
\equiv {\hat k}^{{\hat \mu}_6}{\hat {\tilde C}}_{{\hat \mu}_1\dots
{\hat \mu}_6}$, and ${\hat b}^{(1)}$ is a 1-form worldvolume field
which describes the coupling to an M2-brane 
wrapped on the Taub-NUT direction. The same coupling appears in the
effective action of the {\bsb} pair, where now 
${\hat b}^{(1)}={\hat b}^{(1)}_1-{\hat b}^{(1)}_2$, with 
${\hat b}^{(1)}_{1,2}$ the corresponding vector field in the worldvolume
of each monopole.
After condensation of the
tachyonic mode of the M2-brane, the integration of the localised flux  
$d{\hat b}^{(1)}$ on a transverse $R^2$ gives

\begin{equation}
\int_{R^{4+1}} i_{\hat k}{\hat {\tilde C}}
\end{equation}

\noindent i.e. an M5-brane soliton with one worldvolume
direction wrapped around the Taub-NUT direction. Therefore we can describe
a (wrapped) M5-brane soliton, as the condensation of an
M2-brane stretched between the Kaluza-Klein anti Kaluza-Klein 
monopole pair.

We can now analyse the different possible processes in the type IIA theory
to which this process gives rise.
Dimensionally reducing along the Taub-NUT direction of the monopole
we can describe a D4-brane through the condensation of an open
string stretched between a D6, anti-D6 pair \cite{sena}. 
This is described by the
coupling:

\begin{equation}
\int_{R^{6+1}}C^{(5)}\wedge db^{(1)}
\end{equation}

\noindent which is straightforwardly obtained by reducing the coupling
(\ref{M2deM5}) describing the creation of the solitonic 
(wrapped) M5-brane. 

The reduction along a worldvolume direction of the monopole gives as
one of the possible configurations a solitonic NS5-brane, 
obtained after the
condensation of an open string stretched between a Type IIA pair
of Kaluza-Klein anti-Kaluza-Klein monopoles.
The worldvolume reduction of (\ref{M2deM5}) gives:

\begin{equation}
\int_{R^{5+1}}i_k C^{(5)}\wedge db^{(1)}+\int_{R^{5+1}}i_k B^{(6)}\wedge
db^{(0)}
\end{equation}

\noindent where $b^{(0)}$ arises as the component of ${\hat b}^{(1)}$
along the worldvolume direction that is being reduced. These terms
describe two processes: One in which a (wrapped) D4-brane is created
after condensation of a (wrapped) D2-brane stretched between two Type
IIA monopoles, described by the first
term, and one in which a (wrapped) NS5-brane is created after the
condensation of a (wrapped) open string.
This is, to our knowledge, the first example in which
a NS5-brane has been described through a brane anti-brane pair
annihilation. 

There is another configuration giving rise to a solitonic NS5-brane,
though it occurs after the annihilation of a pair of so-called
exotic branes, i.e. branes not predicted by the analysis of the
spacetime supersymmetry algebra.
This process is obtained after the reduction of the M2-brane
stretched between the KK, anti-KK pair along a direction transverse
to the monopoles, but different from the Taub-NUT direction. 
This reduction of the M-theory Kaluza-Klein monopole gives rise to
a Kaluza-Klein type of solution in ten dimensions whose transverse
space is not a four dimensional euclidean Taub-NUT space, as for the
conventional Kaluza-Klein monopole, but three dimensional. Therefore
this solution is not asymptotically flat but logarithmically divergent.
Moreover it is not predicted by
the analysis of the Type IIA spacetime supersymmetry algebra (see 
\cite{EL} for a possible explanation of this fact). We will however
discuss briefly this type of configurations because they give rise to
interesting descriptions of NS-NS branes in terms of 
brane anti-brane annihilation.
We will denote the brane obtained through this 
reduction of the M-theory KK-monopole as a KK6-brane\footnote{The Type 
IIA Kaluza-Klein monopole would then be denoted as a KK5-brane.}.
The coupling in the KK6-brane
worldvolume action reads:

\begin{equation}
\int_{R^{6+1}}i_k B^{(6)}\wedge db^{(1)}\, .
\end{equation}

\noindent This describes a (wrapped) NS5-brane after condensation
of a (wrapped) D2-brane stretched between a KK6, anti-KK6 pair of branes.

We now turn to the ``dual'' process in the {\bsb} system  , i.e.  
obtaining an M2-brane soliton after the condensation of an M5-brane. 
A wrapped M5-brane is coupled in the worldvolume of
a higher dimensional brane through a 4-form field, which in the 7 
dimensional worldvolume of the monopole is dual to the vector field
${\hat b}^{(1)}$ describing the coupling of a wrapped M2-brane. Given a 
dual pair of worldvolume fields only one can couple at the same time
in the worldvolume effective action, given that both of them carry the
same number of degrees of freedom. Therefore we need to dualise the
vector field ${\hat b}^{(1)}$ in the Kaluza-Klein monopole effective
action. First, one adds a term

\begin{equation}
\label{dual}
\int_{R^{6+1}}d{\hat b}^{(1)}\wedge d{\hat b}^{(4)}
\end{equation}

\noindent to the action, and then integrates ${\hat b}^{(1)}$ from its
equation of motion\footnote{This step is normally quite involved due
to the complicated form of the Born-Infeld part.}.
Since ${\hat b}^{(1)}$ couples 
through its gauge invariant field strength:

\begin{equation}
{\hat F}^{(2)}=d{\hat b}^{(1)}+ (i_{\hat k}{\hat C})\, ,
\end{equation}

\noindent we can write (\ref{dual}) as

\begin{equation}
\int_{R^{6+1}}({\hat F}^{(2)}-(i_{\hat k}{\hat C}))\wedge
d{\hat b}^{(4)}\, ,
\end{equation}

\noindent from where a term 

\begin{equation}
\label{dualb4}
\int_{R^{6+1}}(i_{\hat k}{\hat C})\wedge d{\hat b}^{(4)}
\end{equation}

\noindent is already known to be coupled to the dual action without
the need to eliminate explicitly ${\hat F}^{(2)}$ from its equation
of motion. This term describes an M2-brane soliton. Indeed, the
condensation of the wrapped M5-brane, accompanied by a localised
$d{\hat b}^{(4)}$-flux over a transverse $R^5$, gives a coupling:

\begin{equation}
\int_{R^{1+1}}(i_{\hat k}{\hat C})\, ,
\end{equation}

\noindent which is the way the eleven dimensional 3-form couples 
to a wrapped M2-brane. Therefore, this is the soliton that is
produced after the condensation.

The creation of a fundamental string in the Type IIA theory is now
described as the reduction of the stretched M5-brane between the
Kaluza-Klein anti Kaluza-Klein pair over the Taub-NUT direction of the
monopole. This gives a D4-brane stretched between a D6, anti-D6
pair. The reduction of the coupling (\ref{dualb4}) gives:

\begin{equation}
\int_{R^{6+1}} B^{(2)}\wedge db^{(4)}\, .
\end{equation}

\noindent Therefore, we find:

\begin{equation}
\int_{R^{1+1}}B^{(2)}
\end{equation}

\noindent after the integration of the 4-form, describing a solitonic
fundamental string \cite{Yi}. 

Three other possible brane anti-brane annihilation processes can be
deduced in Type IIA 
from the M-theory Kaluza-Klein anti-Kaluza-Klein annihilation
that we have just discussed.
If we reduce this process along a worldvolume direction of the
Kaluza-Klein monopole we find the following couplings:

\begin{equation}
\label{KK5s}
\int_{R^{5+1}}i_k B^{(2)}\wedge db^{(4)}+\int_{R^{5+1}}i_k C^{(3)}
\wedge db^{(3)}
\end{equation}

\noindent where:

\begin{equation}
{\hat b}^{(4)}_{\mu_1\dots\mu_4}=b^{(4)}_{\mu_1\dots\mu_4}\, ,\qquad
{\hat b}^{(4)}_{\mu_1\dots\mu_3 6}=b^{(3)}_{\mu_1\dots\mu_3}\, ,\qquad
\mu=0,1,\dots,5\, .
\end{equation}

\noindent The first term in (\ref{KK5s}) describes a fundamental string
created after the condensation of a NS5-brane stretched between
a pair of Type IIA Kaluza-Klein monopole anti-monopole. 
Both the fundamental string and
the NS5-brane are wrapped on the Taub-NUT direction of the 
monopole. On the other hand, the second term represents a wrapped
D2-brane arising after the condensation of a wrapped D4-brane
stretched between the two monopoles. 
The third case is obtained reducing (\ref{dualb4}) along a 
direction transverse to the monopole but different from its 
Taub-NUT direction.
We obtain the following coupling in the seven dimensional worldvolume
of a KK6-brane:

\begin{equation}
\int_{R^{6+1}}i_k C^{(3)}\wedge db^{(4)}\, .
\end{equation}

\noindent This describes a D2-brane wrapped around the NUT direction of
the KK6-brane, and it is obtained after the condensation of a (wrapped)
NS5-brane stretched between the brane anti-brane pair.

\section{The {\bsc} system}

In this section we study the possible processes of creation
of branes after tachyonic condensation in the {\bsc} system. 
We identify the coupling in the M9-brane  effective action responsible 
for this condensation and analyse the related processes of creation of  
branes in type IIA.

For this purpose it is important to recall that the M9-brane 
contains a gauged direction in its worldvolume \cite{BvdS}, 
such that the field content is that of the nine dimensional vector
multiplet. Reduction along this direction
gives the D8-brane effective action, whereas the reduction along a 
different worldvolume direction gives the KK8-brane, another
so-called exotic brane of the
Type IIA theory, in the sense that it is not predicted by the analysis
of the Type IIA spacetime supersymmetry algebra. This brane contains as
well a gauged direction in its worldvolume, inherited from that of
the M9-brane, and has been studied
in connection with the 7-brane of the Type IIB theory in
\cite{EL}, where its worldvolume effective action has been derived. 
The Wess-Zumino term of the M9-brane effective action has not yet 
been constructed in the literature. However, from the effective
action of the KK8-brane we can deduce the presence of the
following term in the M9-brane worldvolume effective action:

\begin{equation}
\label{M9coup}
\int_{R^{8+1}} i_{\hat k}{\hat N}^{(8)}\wedge d{\hat b}^{(1)}\, .
\end{equation}

\noindent Let us stress that the M9-brane is effectively an 8-brane,
given that it has one of its worldvolume directions gauged.
Therefore the integration takes place over a nine dimensional 
space-time.
The contraction of the 8-form potential ${\hat N}^{(8)}$ with the
(worldvolume) Killing direction, denoted by ${\hat k}$,
is the field to which the M-theory Kaluza-Klein
monopole couples minimally\footnote{As we mentioned in the previous
section here ${\hat N}^{(8)}$ is the electric-magnetic dual of the
Killing vector considered as a 1-form.}. ${\hat b}^{(1)}$ is a 1-form
worldvolume field describing a wrapped M2-brane ending on the 
M9-brane \cite{EL2}. 
The reduction of this term along a worldvolume direction 
different from the gauged direction gives the terms:

\begin{equation}
\label{KK8}
\int_{R^{7+1}}i_k N^{(8)}\wedge db^{(0)}+\int_{R^{7+1}}i_k N^{(7)}
\wedge db^{(1)}
\end{equation}

\noindent present in the effective action of the KK8-brane (expression
(4.6) of reference \cite{EL}), and where:

\begin{equation}
{\hat b}^{(1)}_\mu=b^{(1)}_\mu\, ,\qquad
{\hat b}^{(1)}_8=b^{(0)}\, , \mu=0,1,\dots,7\, ,
\end{equation} 

\noindent and $i_{\hat k}{\hat N}^{(8)}$ gives rise to the two fields
$i_k N^{(8)}$, $i_k N^{(7)}$. $i_k N^{(8)}$ is the field that couples
minimally to the KK6-brane that we considered in the previous section,
whereas the ordinary Type IIA Kaluza-Klein monopole is charged with
respect to $i_k N^{(7)}$. 

The coupling (\ref{M9coup}) in the M9-brane worldvolume effective action 
describes the process in which a Kaluza-Klein
monopole is created after the condensation of a, wrapped, M2-brane
stretched between an M9, anti-M9 pair of branes. As in the previous
sections the pair is described by choosing ${\hat b}^{(1)}$ as the
difference of vector fields in each brane.
Indeed, integration over the localised magnetic 
${\hat b}^{(1)}$-flux, associated to the wrapped M2-brane, on
a transverse $R^2$ gives:

\begin{equation}
\int_{R^{6+1}} i_{\hat k}{\hat N}^{(8)}
\end{equation}

\noindent i.e. the field minimally coupled to the Kaluza-Klein
monopole.

Reducing (\ref{M9coup}) along the isometric worldvolume 
direction denoted by ${\hat k}$ we can describe the
process in which a D6-brane is created after the condensation of an open
string stretched between a pair of D8, anti-D8 branes \cite{sena}. 
The resulting coupling in the worldvolume effective action of the D8
brane is given by:

\begin{equation}
\int_{R^{8+1}}C^{(7)}\wedge db^{(1)}
\end{equation}

\noindent where $C^{(7)}$ is the RR 7-form potential of the Type IIA
theory and is obtained through the reduction:

\begin{equation}
i_{\hat k} {\hat N}^{(8)}=C^{(7)}+\dots
\end{equation}

\noindent (see \cite{BEL} for the details of this reduction). Therefore
the condensation of a fundamental string, accompanied by a localised
$db^{(1)}$ flux over a transverse $R^2$, gives a D6-brane soliton:

\begin{equation}
\int_{R^{6+1}}C^{(7)}\, .
\end{equation}

Instead, we can reduce (\ref{M9coup}) along a worldvolume direction of the
M9-brane, in which case we obtain a pair of KK8, anti-KK8 branes
with the following couplings:

\begin{equation}
\int_{R^{7+1}}i_k N^{(7)}\wedge db^{(1)}+\int_{R^{7+1}}i_k N^{(8)}
\wedge db^{(0)}\, ,
\end{equation}

\noindent derived already in (\ref{KK8}). 
The first coupling describes a Type IIA Kaluza-Klein monopole,
obtained after the condensation of a D2-brane. The Taub-NUT direction
of the monopole coincides with the gauged worldvolume direction of the KK8
branes, and the stretched D2-brane is also wrapped around this direction.
The second coupling describes a KK6-brane, obtained
after the condensation of a (wrapped) open string stretched between
the two KK8, anti-KK8 branes. Again the Killing direction of the KK6 and
KK8-branes coincide.
 A last reduction on (\ref{M9coup}) can be performed along the transverse 
direction. Such a reduction on the M9 gives rise to the NS9A-brane
predicted by the Type IIA spacetime supersymmetry algebra \cite{Hull}.
We have thus the possibility of obtaining a KK6 after
the condensation of a (wrapped) D2 stretched between a NS9A, anti-NS9A
pair.

\vspace{0.5cm}

As in the previous section, the process in which a wrapped M2-brane is
created through the condensation of a Kaluza-Klein monopole stretched
between the pair of M9, anti-M9 branes is described by the coupling 
dual to (\ref{M9coup}). The Kaluza-Klein monopole must be coupled to
the (9 dimensional) worldvolume of the M9-brane
through a 6-form worldvolume field ${\hat b}^{(6)}$, which must be
the worldvolume dual of the vector field ${\hat b}^{(1)}$. In the 
dualisation process we find:

\begin{equation}
\int_{R^{8+1}} d{\hat b}^{(1)}\wedge d{\hat b}^{(6)}=
\int_{R^{8+1}} ({\hat F}^{(2)}-(i_{\hat k}{\hat C}))\wedge
d{\hat b}^{(6)}\, ,
\end{equation} 

\noindent from where we already identify a coupling:

\begin{equation}
\label{M9M6}
\int_{R^{8+1}}i_{\hat k}{\hat C}\wedge d{\hat b}^{(6)}
\end{equation}

\noindent in the dual effective action. Integration over a localised
magnetic ${\hat b}^{(6)}$-flux, associated to
the Kaluza-Klein monopole, on a transverse $R^7$ gives:

\begin{equation}
\int_{R^{1+1}}i_{\hat k}{\hat C}
\end{equation}

\noindent describing a wrapped M2-brane soliton.

The reduction of (\ref{M9M6}) along the ${\hat k}$-direction gives:

\begin{equation}
\int_{R^{8+1}}B^{(2)}\wedge db^{(6)}.
\end{equation}

\noindent Here $b^{(6)}$ is associated to a D6-brane, and the
integration of its localised magnetic flux on
a transverse $R^7$ gives:

\begin{equation}
\int_{R^{1+1}}B^{(2)}\, ,
\end{equation}

\noindent describing a fundamental string soliton \cite{Yi}.

As before, we can also analyse the process obtained by reducing 
(\ref{M9M6}) along a worldvolume direction. We obtain:

\begin{equation}
\int_{R^{7+1}}i_k C^{(3)}\wedge db^{(5)}+\int_{R^{7+1}}
i_k B^{(2)}\wedge db^{(6)}\, ,
\end{equation}

\noindent where $b^{(5)}$ arises from the reduction of ${\hat b}^{(6)}$
along the worldvolume direction. The first term describes a (wrapped)
D2-brane, occuring after the condensation of a KK5 monopole
whose Killing direction coincides with the Killing direction of the
KK8, anti-KK8 branes obtained after the reduction.
The second term describes a (wrapped) fundamental string,
realised in this case after the condensation of a KK6-brane stretched
between the pair of KK8, anti-KK8 branes. Finally, reducing 
(\ref{M9M6}) along the
transverse direction leads to a process where a (wrapped) D2-brane
is produced after the condensation of a KK6-brane stretched between a pair 
of NS9A, anti-NS9A branes.

\section{Brane-antibrane systems in M-theory involving M-waves}

In this section we analyse, among other processes, 
the brane-antibrane system in M-theory
giving rise to the process in which a fundamental string stretched
between a D2, anti-D2 pair condenses and a solitonic D0-brane is produced.
This process is described in M-theory by a wrapped M2-brane 
stretched between a pair M2, anti-M2. When the tachyonic mode of the
wrapped M2-brane condenses an M-wave soliton is produced. In order to
identify the Wess-Zumino coupling in the worldvolume effective action of
the M2, anti-M2 pair of branes that is responsible for this process 
let us first analyse its dual configuration,
i.e. that in which an M-wave is ``stretched" between the brane
anti-brane pair and a wrapped M2-brane is created after its tachyonic
mode condenses. 

The so-called M-wave is a pp-wave in M-theory
carrying momentum along a given direction, which we will denote by $y$.
An M-wave ending on another M-brane is described in the worldvolume
effective action of the latter by its coupling to the embedding
scalar $y$. Therefore in order to have a non-trivial 
condensation of the tachyonic mode of the M-wave
this direction must be topologically non-trivial. Indeed, 
the coupling in the worldvolume effective action of the 
M2, anti-M2 pair of branes that is responsible for the condensation
of the tachyonic mode of the M-wave is given by:

\begin{equation}
\label{original}
\int_{R^{2+1}} i_{\hat h}{\hat C}\wedge dy\, ,
\end{equation}

\noindent where ${\hat h}$ denotes a Killing vector in the 
$y$-direction. 
Integration over a localised magnetic $dy$-flux gives a coupling:

\begin{equation}
\int_{R^{1+1}}(i_{\hat h}{\hat C})\, ,
\end{equation}

\noindent describing an M2-brane wrapped on the $y$-direction.
The dual process is now described by dualising this coupling. For this
we need to recall that the field strength associated to the $y$-field,
which appears in the worldvolume effective action of the M-theory
pp-wave \cite{BT}, 
is ${\hat F}^{(1)}=dy+{\hat h}^{-2}{\hat h}_\mu dx^\mu$, 
where $\mu$ runs
over all directions but the $y$ direction. Therefore, we find a coupling:

\begin{equation}
\label{dualbis}
\int_{R^{2+1}}{\hat h}^{-2}{\hat h}_\mu dx^\mu \wedge d{\hat b}^{(1)}\, ,
\end{equation}

\noindent ${\hat b}^{(1)}$ being the worldvolume dual of the scalar $y$.
Integrating now over a localised ${\hat b}^{(1)}$-flux we end up with
the coupling:

\begin{equation}
\int_R {\hat h}^{-2}{\hat h}_\mu dx^\mu
\end{equation}

\noindent which is the field minimally coupled to the M-wave moving
in the ${\hat h}$ direction. Therefore condensation of the wrapped
M2-brane coupled to ${\hat b}^{(1)}$ in the worldvolume of the pair
produces an M-wave moving in the direction on which the stretched
M2-brane is wrapped.

We can now analyse which are the Type IIA brane anti-brane annihilation 
processes to which these two systems give rise. 
It is particularly interesting
to consider the reduction along the $y$-direction. The coupling
(\ref{original}) gives rise to:

\begin{equation}
\int_{R^{2+1}}B^{(2)}\wedge dy\, .
\end{equation}

\noindent Therefore it describes a fundamental string, through the
condensation of a D0-brane stretched between the D2, anti-D2 pair of
branes obtained after the reduction. On the other hand, the coupling
(\ref{dualbis}) gives:

\begin{equation}
\int_{R^{2+1}}C^{(1)}\wedge db^{(1)}
\end{equation}

\noindent which in turn describes a solitonic D0-brane after condensation
of a fundamental string. Therefore we can conclude that the M2, anti-M2
systems that we have discussed are the origin in M-theory of both the
D0-brane creation studied by Sen \cite{sena} and the creation of a 
fundamental string
in a D2, anti-D2 system after D0-brane condensation \cite{Yi}.

Reduction along a direction different from $y$ gives rise to two
interesting processes which, as we will see, generalise to all branes in
Type II theories. From worldvolume reduction we obtain a process
in which a wrapped fundamental string stretched\footnote{
The string is in fact not really properly stretched between
 the pair being wrapped in the $y$--direction.}
between a pair F1, anti-F1 gives rise to a pp-wave moving 
in the direction on which the stretched
string is wrapped. Of course we also find the dual to this process,
i.e. a solitonic wrapped fundamental string emerging after the condensation
of a pp-wave in the same brane configuration. Similarly,
after doing a direct dimensional reduction we obtain the same 
type of configurations
but for D2-branes. A wrapped D2-brane is created after a pp-wave condenses
in a D2, anti-D2 pair, and a pp-wave can also be created if instead a 
wrapped D2-brane condenses. 

An M-wave, being coupled in the worldvolume of an M-brane
through the worldvolume scalar labelling the direction in which
it propagates, can end on any of the branes of M-theory. Therefore 
we can analyse the same two processes that we have studied in the
{\bsd} case on M5, MKK or M9 
brane anti-brane systems. We are not going to repeat in detail the 
corresponding analysis of the Wess-Zumino terms responsible for these
processes, since the reasoning goes straightforwardly as for the M2 system.
Let us just mention that, together with the process in which 
any wrapped Type IIA p-brane\footnote{Also NS-NS ones, like
the fundamental string that we have just considered.}
stretched between a p-brane anti p-brane pair gives rise to a pp-wave
soliton, and its dual (wrapped p-brane creation through condensation of
a pp-wave), we find the following processes:

$$({\rm NS5},\overline{\mbox{NS5}}; {\rm D4}\rightarrow {\rm D0})$$
$$({\rm NS5},\overline{\mbox{NS5}}; {\rm D0}\rightarrow {\rm D4})$$ 
$$({\rm KK6},\overline{\mbox{KK6}}; {\rm KK5}\rightarrow {\rm D0})$$
$$({\rm KK6},\overline{\mbox{KK6}}; {\rm D0}\rightarrow {\rm KK5})$$
$$({\rm NS9},\overline{\mbox{NS9}}; {\rm KK8}\rightarrow {\rm D0})$$
$$({\rm NS9},\overline{\mbox{NS9}}; {\rm D0}\rightarrow {\rm KK8})$$

\noindent where we use a simplified notation to indicate that the condensation
of a brane (third column) stretched between the brane-antibrane
pair given by the first and second columns gives rise to a certain
brane soliton, specified by the fourth column. 
Most of these processes involve exotic branes, but it is particularly
interesting to see that a D0-brane can be realised as a solitonic
configuration in an NS5, anti-NS5 annihilation process. This represents a 
novel way of realising this brane soliton other than through,
the more conventional, D1, anti-D1 annihilation process \cite{sena}.

\section{Type IIB branes from brane-antibrane systems}

In the previous sections we have derived all the Type IIA
branes that can possibly occur as solitons after the condensation 
of a tachyonic
brane stretched between a pair brane-antibrane. This derivation was
performed by direct reduction from M-theory, where the Wess-Zumino
term in the worldvolume effective action of the brane-antibrane pair
responsible for the process was identified.
It is now straightforward
to derive the T-dual couplings responsible for brane creation in a 
brane-antibrane system in the Type IIB theory. 

Together with the realisation of Dp-branes as bound states of D(p+2), 
anti D(p+2) branes \cite{sena} we find the corresponding dual processes,  
in which a BPS fundamental string is described as a bound state of 
D(p+2), anti D(p+2) branes after the condensation of a tachyonic 
Dp-brane stretched between them. T-duality allows to identify explicitly
the couplings in the D(p+2)-brane effective action that are responsible
for these processes:

\begin{equation}
\int_{R^{p+3}}C^{(p+1)}\wedge db^{(1)}
\end{equation}

\noindent describes Dp-brane creation after condensation of an F1,
described by $b^{(1)}$, and:

\begin{equation}
\int_{R^{p+3}} B^{(2)}\wedge db^{(p)}
\end{equation}

\noindent describes F1 creation after condensation of the p-brane
described by $b^{(p)}$, worldvolume dual of $b^{(1)}$.

Type IIB branes are
organised as singlets or doublets under the Type IIB 
SL(2,Z) duality group.
Therefore we expect to find a spectrum of brane solitons in
brane-antibrane systems that respects this symmetry. 
The D5 brane forms an SL(2,Z) doublet with the NS5 brane
of the Type IIB theory. Therefore S-duality predicts the
occurrence of a D3-brane soliton as a bound state in a 
NS5, anti-NS5 pair when the condensation of a D1-brane
stretched between the two takes place. 
This process is indeed obtained after
T-dualising two Type IIA configurations. Namely, that in which a D2 
brane stretched between a pair of NS5, anti-NS5 branes condensed to 
give a BPS D2-brane, and that in which the same kind of brane was
stretched between a pair of KK5 monopoles and condensed to give rise
to a D4-brane soliton. Of course the process in which a D1-brane
is created after the condensation of a D3-brane stretched between the
two NS5, anti-NS5 branes is also predicted by T-duality from certain
Type IIA configurations. For these and the following configurations
we omit the explicit Wess-Zumino couplings, since they can be
straightforwardly derived from those in Type IIA found in the
previous sections.
The situation with the D9-brane is similar to that with the D5-brane,
in the sense that it forms a doublet with an NS9-brane in the
Type IIB theory. We have indeed obtained after T-duality the
configurations describing a 7-brane as a bound state NS9, anti-NS9
after a stretched D1-brane condenses, and its dual, namely a D1-brane
realised as a bound state NS9, anti-NS9 through the condensation of a
tachyonic 7-brane.

For the Type IIB {\bsb} system we find that
a pair of Type IIB KK, $\overline{\mbox{KK}}$ monopoles can annihilate
giving rise to the following solitonic branes:
$$({\rm KK5},\overline{\mbox{KK5}}; {\rm D3}\rightarrow {\rm D3})$$
$$({\rm KK5},\overline{\mbox{KK5}}; {\rm D1}\rightarrow {\rm D5})$$
$$({\rm KK5},\overline{\mbox{KK5}}; {\rm F1}\rightarrow {\rm NS5})$$
$$({\rm KK5},\overline{\mbox{KK5}}; {\rm D5}\rightarrow {\rm D1})$$
$$({\rm KK5},\overline{\mbox{KK5}}; {\rm NS5}\rightarrow {\rm F1})$$
where we use the same simplified notation as in the previous section.
We see from these configurations that the creation of both a brane
and its S-dual is possible through annihilation of a pair of monopoles,
which agrees with the fact that this brane is selfdual under
S-duality, so all pairs of S-dual processes should be allowed.

Concerning 7-branes some remarks are in order. 
T-duality of the Type IIA KK6 and KK8-branes predicts a
7-brane in the Type IIB theory which is connected by S-duality with
the D7-brane, for what we will denote it as an NS7-brane.
This is the 7-brane that appears for instance in the
processes involving NS9, anti-NS9 annihilation.
The existence of these
two different effective actions describing 7-branes in the Type IIB theory, 
where the analysis of
the spacetime supersymmetry algebra predicts a single 7-brane, was
discussed in \cite{EL}, where it was argued that the two worldvolume
effective actions are indeed necessary in order to describe a single
nonperturbative 7-brane in the weak and strong coupling regimes.
Consistently with this picture, we find after T-duality one configuration
in which an NS5-brane is created when a D1-brane stretched between a pair
of NS7, anti-NS7 branes condenses, and the corresponding dual process, i.e.
that in which a solitonic D1-brane emerges after condensation of an
NS5-brane. These processes are S-dual to the more standard D5, F1 
creation \cite{sena} \cite{Yi} through D7, anti-D7 annihilation.  

Finally, T-duality on the Type IIA configurations involving pp-waves
predicts the analogous kind of configurations in the Type IIB theory.
Namely, solitonic wrapped p-branes after p-brane anti p-brane annihilation
through condensation of a pp-wave, and solitonic pp-waves after 
condensation of a wrapped p-brane in the same system.

\section{Discussion}

In this paper, extending preceeding considerations \cite{Yi}, 
we have classified the branes of the M and Type II theories
which can be realised in a consistent way with respect to the 
structure of the theory as bound states of systems 
brane-antibrane, after condensation of the tachyon mode of open branes 
stretched between the pair. 
We have achieved this classification by studying the 
Wess-Zumino terms in the worldvolume effective actions of the branes
of M-theory and their reductions.

We have shown that it is possible to give an eleven dimensional 
description to the creation of a fundamental string from the
annihilation of a pair of Dp, anti-Dp branes with p=2,6,8. 
As for the case p=4 \cite{Yi}, the fundamental string 
is created by the condensation of a D(p-2)-brane stretched between the
pair Dp, anti-Dp.
We have identified the term in the Wess-Zumino action of the pair
brane-antibrane in M-theory responsible for this creation. This term is
obtained generically by finding the worldvolume dual of
the coupling describing a stretched
M2-brane between the brane anti-brane pair, which in turn describes an
extended worldvolume soliton coming from the condensation of the M2-brane.
The case p=4 is especially simple in this sense.
Its description in M-theory is in terms of an M2-brane 
stretched between a pair of M5, anti-M5 branes, giving rise to a
solitonic M2-brane. This process is self-dual, in the sense that the
dual coupling in the six dimensional worldvolume of the M5-brane 
describes again a solitonic M2-brane, whereas this is not the case for the
M-theory origin of the p=2,6,8 cases.
Therefore, in these cases
the reduction to the Type IIA theory gives many different 
realisations of BPS objects as bound states of brane anti-brane pairs,
which by T-duality give in turn rise to many configurations
in the Type IIB theory.
As an interesting realisation we have found that the NS5-brane can 
originate from certain brane anti-brane annihilations, both in the
Type IIA and Type IIB theories.

We can conclude in general that the BPS branes that can be realised
as solitons in a given p-brane anti p-brane configuration are determined
by the possible branes that can end on the p-brane, and whose 
tachyonic mode condenses giving rise to the soliton configuration. 
These branes are
easily predicted by looking at the p-brane worldvolume effective action
and identifying the different worldvolume q-forms that couple in it.
For instance the Type IIB Kaluza-Klein monopole anti-monopole pair
gives rise to so many different solitonic configurations because there
are two scalars and one 2-form coupled in the worldvolume effective
action of the monopole \cite{EJL}, allowing for wrapped F1, D1 and 
D3-branes ending on it, together with
the branes described by their worldvolume duals, i.e. D5 and NS5
branes.

Finally, it is worth emphasizing that a lot of progress remains to
be done in order to reach a  quantitative understanding of the dynamics
of the possible processes discussed in this paper. One can furthermore 
hope  that there exists a mathematical connection with some generalisation
of K-theory \cite{kteo,hora2} where these processes could fit in.

\subsection*{Acknowledgements}

L.~H. would like to acknowledge the support of the European Commission 
TMR programme grant ERBFMBICT-98-2872.

\end{document}